\documentclass[conference]{IEEEtran}

\IEEEoverridecommandlockouts %To add sponsors and index terms

\usepackage{cite}
\usepackage[cmex10]{amsmath}

\ifCLASSINFOpdf
   \usepackage[pdftex]{graphicx}
   \graphicspath{{./Images/}}
   \DeclareGraphicsExtensions{.pdf,.jpeg,.png}
\else

\fi
\usepackage{textcomp}
\usepackage{url,amsfonts}
\usepackage[bookmarks=false,hidelinks]{hyperref}
\usepackage{multirow}
\usepackage{booktabs,colortbl}
\usepackage{todonotes}
\usepackage{cleveref}

\usepackage{tikz}
\usepackage{textcomp}
\usepackage{hyperref}
\usepackage{lipsum}

\newcommand\copyrighttext{%
    \footnotesize \textcopyright 2016 IEEE. Personal use of this material is permitted.
    Permission from IEEE must be obtained for all other uses, in any current or future 
    media, including reprinting/republishing this material for advertising or promotional 
    purposes, creating new collective works, for resale or redistribution to servers or 
    lists, or reuse of any copyrighted component of this work in other works. }
\newcommand\copyrightnotice{%
    \begin{tikzpicture}[remember picture,overlay]
    \node[anchor=south,yshift=10pt] at (current page.south) {\fbox{\parbox{\dimexpr\textwidth-\fboxsep-\fboxrule\relax}{\copyrighttext}}};
    \end{tikzpicture}%
}

\begin{document}
\title{
% Prediction of Sensor Signals in Smart Energy Systems for Monitoring and Anomaly Detection}
%
%Dynamic Programming for Smart Water Heater Control
%Optimal Water Heater Control for Smart Home Applications
Optimal Water Heater Control in Smart Home Environments
}
\author{%
\IEEEauthorblockN{Christoph Passenberg, Dominik Meyer, Johannes Feldmaier, Hao Shen}
\IEEEauthorblockA{Department of Electrical and Computer Engineering,
Technische Universit\"at M\"unchen, Germany \vspace*{-2em}
}
% \thanks{Identify applicable sponsor/s here. \emph{(sponsors)}}%
}
\maketitle
\copyrightnotice
%ABSTRACT
\begin{abstract}
% In this paper we investigate the problem of optimal warm water heater control in a smart home. 
In this work, we develop an optimal water heater control method for a smart 
home environment.
It is important to notice that unlike battery storage systems, energy flow 
in a water heater control system is not reversible.
In order to increase the eigen consumption of photovoltaic energy, i.e. direct consumption of generated energy in the house, 
we propose to employ a dynamic programming approach to optimize heating 
schedules using forecasted consumption and weather data . % and exploit cheap 
Simulation results demonstrate the capability of our proposed system in 
reducing the overall energy cost while maintaining residents' comfort. 
% by increasing the eigen consumption of photovoltaic energy.
%Our system is able to reduce the amount paid to the utility company and to increase
%the immediate consumption of photovoltaic energy.
%
\end{abstract}
%
%INDEX TERMS
\begin{IEEEkeywords}
Dynamic programming, energy cost minimization, smart energy systems, thermal energy storage.
% eigen consumption maximization.
% condition monitoring; fault diagnosis; maintenance engineering; stability monitoring; 
% Decision support systems;Forecasting;Load modeling;Predictive models;State estimation;Weight
% The author shall provide up to 4 keywords (in alphabetical order) to help identify the 
% major topics of the paper. The thesaurus of IEEE indexing keywords is posted at \linebreak \url{http://www.ieee.org/organizations/pubs/ani_prod/keywrd98.txt}\\
%demand side management;electric power generation;energy management systems;power distribution;demand response;demand side management;distributed energy resources;energy distribution;energy generation;energy management;intelligent energy systems;smart loads;sophisticated real-time control;Automation;IEC standards;Load management;Load modeling;Power generation;Pricing;Spinning;Building automation;IEC 61850;demand response;demand side management (DSM);energy efficiency;energy management;load management;peak shaving;smart grids;domestic appliances;power consumption;power generation scheduling;renewable energy sources;smart power grids;HESS;complex scheduling problems;energy cost minimization;energy systems;hybrid electrical storage systems;price-control;residential appliances;residential energy consumption;residential load scheduling;residential renewable resources;smart grids;Batteries;Home appliances;Job shop scheduling;Linear programming;Photovoltaic systems;Schedules;Hybrid Energy Storage Systems;Load Scheduling;Online controllers.
\end{IEEEkeywords}
\section{Introduction}
In modern smart homes, energy produced by a local ge\-ne\-rator, e.g. photovoltaic
(PV) systems, is usually fed into the supply grid directly.
This is due to the domestic
demand-supply mismatch and the lack of
efficient and adequately sized energy storage systems which could store the
generated energy until the residents need it. The temporal mismatch between
electricity production and consumption occurs as the PV systems generate energy
during the day when the residents are mostly leaving the house for work.
Additionally, a big amount of the daily energy consumption occurs during late
hours for the preparation of dinners or for lighting during night hours.
However, feeding the energy directly to the grid increases the difficulty in maintaining its
stability and is economically 
unfavorable for the home owner, since the price that is paid for selling energy to
the grid is usually lower than for buying energy from it. 
% Most of the times, the overall energy
% demand in the grid is low at the same times as for the individual homes. Maintaining grid
% stability therefore becomes more and more difficult for the energy operator. Furthermore, it
% is not optimal to sell all generated power since prices for selling are usually lower than
% for buying energy back from the grid operator. 
%

A common approach to solve the problem of temporal mismatch between generated and
consumed energy is to schedule the usage of electrical devices in the household, cf. \cite{Brunelli2014}, \cite{Galvan2014}. Unfortunately, due to the scheduling
of the household appliances, the residents' comfort might be reduced. Therefore, the
demand-side management has to cope with the trade-off between the two goals of
optimizing the
energy consumption and ensuring the availability of the electrical devices at
all time. As this
trade-off prevents finding a convenient solution for both sides, it is
often still necessary to use an energy storage device.

Electrical energy storage systems like batteries
could solve the problem of the temporal mismatch in domestic energy production and
demand.
Optimal battery charging strategies have been extensively studied, cf.
\cite{Wei2014},\cite{Squartini2_2013},\cite{Welch2007}. However, battery systems
are still very expensive and limited in storage capacity. 

In this work, we investigate the possibility of using a warm water tank as energy
storage. The major difference to batteries is that the energy cannot be converted
back into 
% electrical
% energy 
electricity
easily and efficiently, once it is stored in thermal form. 
Thus, other than the classic problem of optimal battery control, our approach
considers both optimal water heating schedules and a minimization of energy costs
in the smart home.
% , distinguishes itself significantly from classical energy storage system control problems.
However, both control optimization problems rely on predicted data for energy
production, demand and prices. Therefore, adequate optimization algorithms have to
be able to cope with uncertainties in forecasts. We apply a Dynamic Programming (DP) 
algorithm \cite{bertsekas1995dynamic} for smart water heater control in our
simulated household. 

\begin{figure}[t]
\centering
\includegraphics[width=0.46\textwidth]{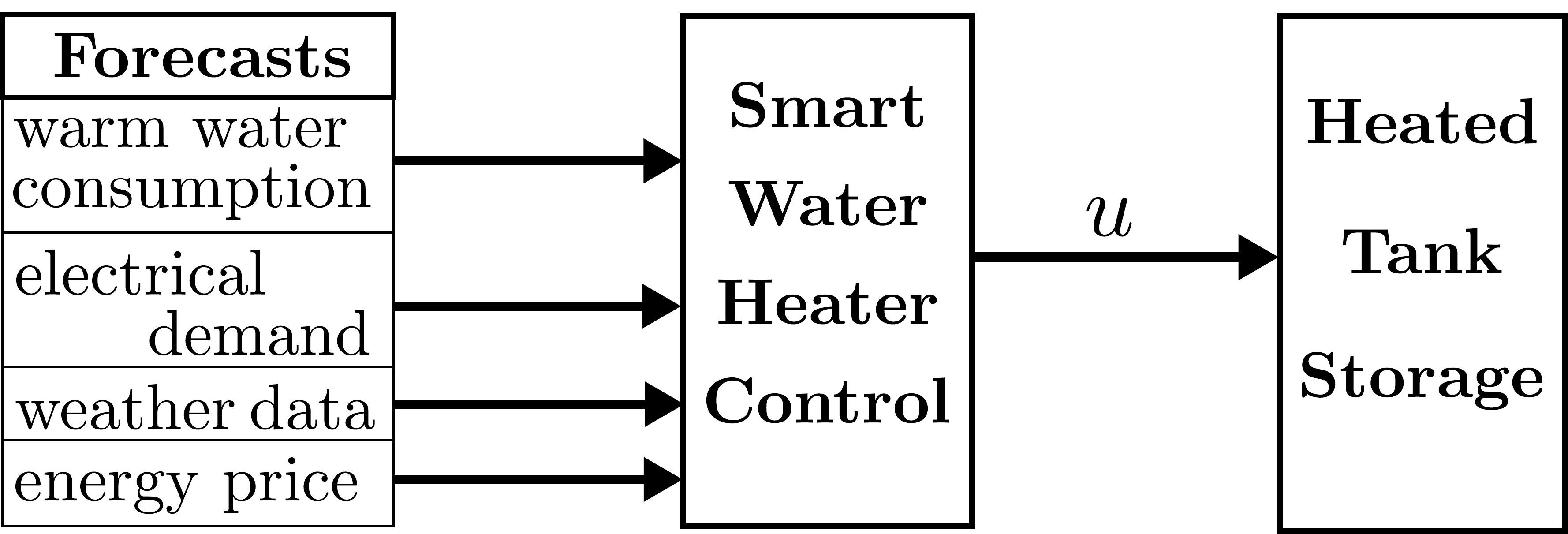}
%\vspace{-0.5em}
\caption{Information flow in the smart water heater control.}
\label{fig:info_flux}
\vspace{-0.6cm}
\end{figure}

In our setting, the goal of the optimization process is to minimize the
energy costs paid to the utility company by finding the optimal control in every
possible
state, while considering uncertainties of the system. The controlled variable is
the power
input to the warm water tank. In order to minimize the future costs, predictions of
warm
water consumption, electrical demand, energy prices and weather data are required
as
illustrated in Fig.~\ref{fig:info_flux}. 
%\vspace{-0.5cm}
%
The reliability of the forecasts influences the 
quality of the
optimal control solution significantly. Hence, it is crucial to have reliable data
in classical
control systems. Existing works such as \cite{Riffonneau2011} 
%neglect 
do not explicitly consider the occurring uncertainties.
% However, our algorithm is able to adaptively cope with the remaining uncertainty
% in the predictions, which is directly considered in the optimization process. 
%However, our algorithm adaptively considers the remaining uncertainty in the predictions.
However, they are adaptively considered in our stochastic DP algorithm.
Another 
distinguishing 
%advantage 
feature of our implementation
%to state of the arts
compared to the state of the arts, e.g. neural
networks \cite{fuselli2012}, is
that we do not need to adapt algorithmic parameters like learning rates. 
% Therefore, no
% parameter tuning is necessary. Instead 
Only parameters of the simulation models for the
warm water tank and the PV-system have to be determined once for each household. 
As our system just employs forecasted data, it is hence independent from the choice
of
prediction model and consequently extendable to use any other prediction method.
% As we just
% use forecasted data, our system is 
% independent of the
% prediction model that is employed to generate the forecasts. 
Additionally, the simulation can be run for different energy price schemes, e.g.
constant and real time pricing. For the latter,
an additional price forecast can be easily integrated in the optimization process.

The rest of the paper is organized as follows:
%The remaining parts of the paper are organized 
%\tcb{as follows:}
%in the following way: 
In Section~\ref{sec:02}, the components of the smart home are presented. Then, a mathematical
description of the optimization problem is stated and the algorithm that
optimizes the control of the smart water tank in order to minimize the energy
costs, is introduced in Section~\ref{sec:03}. In Section~\ref{sec:setting} the experimental setting with its
parameters is presented and the compared methods are shortly explained.
Subsequently, in Section~\ref{sec:result} the results of the simulation runs are
visualized and discussed. Section~\ref{sec:con} contains the conclusions.

\section{Model of simulated Smart Home}
\label{sec:02}
\begin{figure}[t]
\centering
\includegraphics[width=0.46\textwidth]{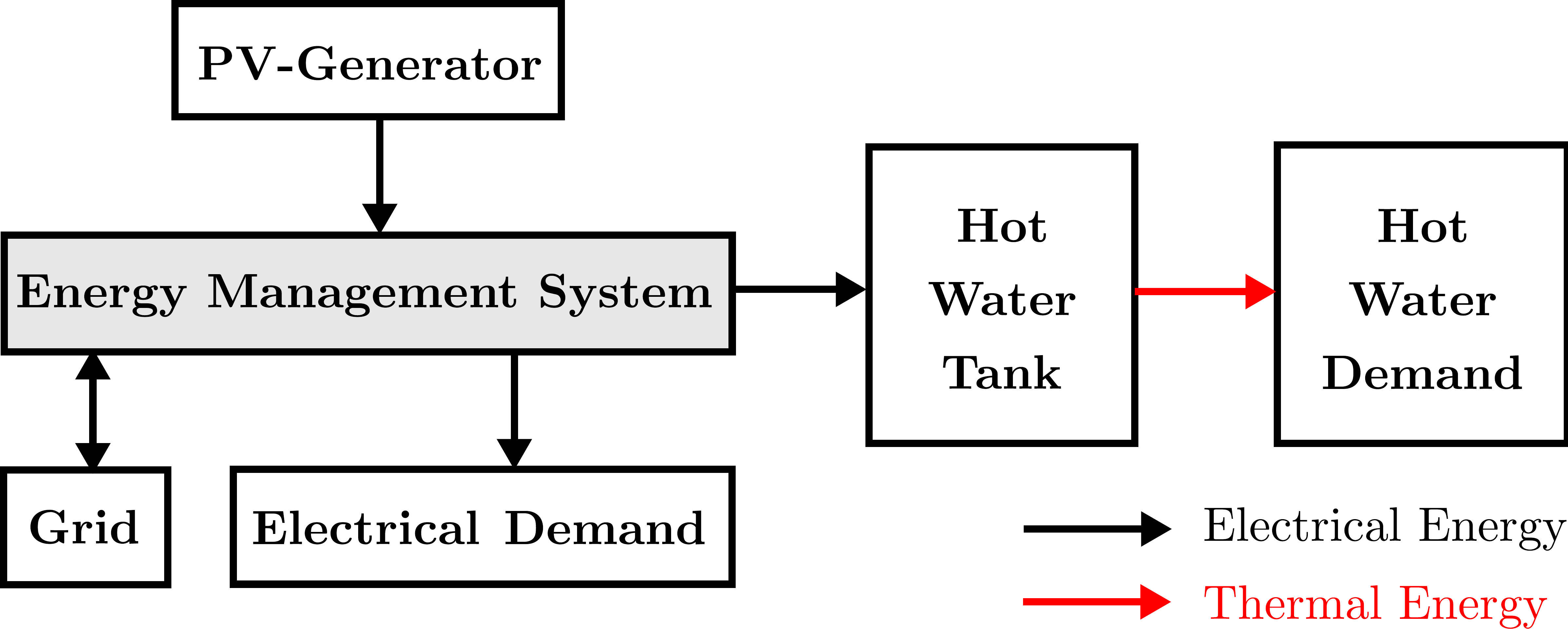}
%\vspace{-0.5em}
\caption{Energy flow in the smart home}
\label{fig:energy_flow}
\vspace{-0.5cm}
\end{figure}

Fig.~\ref{fig:energy_flow} depicts an overview of the simulated smart home with its different 
components and the energy flow between them is depicted.
%In this section the simulated smart home with its different \tcp{components} is presented. An overview of its components and the energy flow of the system is given in Fig.~\ref{fig:energy_flow}. 
In this model, the smart home contains an energy management system (EMS), whose task 
is to distribute the produced energy from the PV-generator 
such that the electrical demand of the residents is always covered and smart water heater
control is enforced. If the PV-generator does not deliver enough energy to meet the demand,
the EMS is able to 
%extract 
utilize energy from the grid, or otherwise to feed energy back into the supply network. In
the simulation, there are no restrictions concerning the power exchange with the grid. Thus,
it is assumed that it is always possible to exchange an arbitrary amount of energy 
with the grid supply and that no stability issues occur.

In our simulation, we implement two common price
schemes for demonstration. 
%Two major different price schemes exist, which can be applied in the simulation. 
One are constant prices for selling and buying energy,
and the other is a real-time price scheme at which the resident operates with the 
current market prices. 

The electricity demand, which has to be satisfied at all time, 
%arises 
manifests due to the usage of electrical devices within the smart home. Their
utilization cannot be controlled to avoid a reduction of the residents' comfort.
Since a hot water tank serves as thermal energy storage in our current setting, excessively produced electric energy of the PV-generator 
can be stored for 
%the usage of the residents. 
later use by the residents as thermal energy.
%However, there is only the feasibility to use the stored energy in form of hot water as
%it is not possible to efficiently convert the thermal energy back into electricity. 
However, it is only possible to store energy by electrically heating up water in the tank, because
in the considered scale, a conversion back to electricity is not efficient and therefore 
not considered further.
%in the form of hot water as it is not possible to efficiently convert the thermal energy back to electricity.}
Similar to the electrical demand, the demand for warm water supply has to be also 
achieved at all time.
%the one for hot water has also to be satisfied at all time. 

In the remainder of this section, the model of the PV-generator and the hot water tank are 
%presented.
introduced and described.
%\vspace{-2cm}
%
\subsection{PV-Generator}
The function that models the power output of the PV-generator dependant on the current
irradiation and outside temperature is specified by
\begin{equation}
\begin{split}
P^{PV}(t)=P_{PV,STC}\frac{I(t)}{I_{STC}}\Bigl(   1-\gamma \Bigr.  &  \Bigl. \left(T_j(t) -T_{j,STC}\right)\Bigr) \\
&\cdot N_{PVs}N_{PVp} \ , 
\label{eq:PV}
\end{split}
\end{equation} 
which is similar to the description in \cite{Riffonneau2011,Brunelli2014}. In Eq.~\eqref{eq:PV} the cell temperature $T_j(t)$ is modeled by
\begin{equation}
T_j(t) = T_{amb}(t)+\frac{I(t)}{I_{NOCT}}\left(NOCT-T_{amb,NOCT}\right) \ ,
\label{eq:Tj}
\end{equation}
%
%similar to} the description  in \cite{Riffonneau2011}
%and has also been used in \cite{Brunelli2014}
%The power which is produced by the PV-generator mainly depends on the current irradiation 
%and outside temperature. This can be modeled by the following equation which is taken from %\cite{Riffonneau2011}
%with the cell temperature
%
where the variables and parameters $P_{PV}$, $P_{PV,STC}$, $I$, $I_{STC}$, $\gamma$, $T_{j,STC}$, 
$N_{PVs}$, $N_{PVp}$, $T_{amb}$, $NOCT$ in  Eq.~\eqref{eq:PV} and Eq.~\eqref{eq:Tj} 
describe the current total power output of all PV-modules, the power output of a PV-module 
under standard test conditions (STC), the current global irradiation on the sloped surface 
of the modules, the irradiation at STC, the power temperature coefficient at the maximum 
power point (MPP), the cell temperature at STC, the number of PV-modules in series, 
the number of PV-modules in parallel, the outside temperature, and the nominal operating 
cell temperature. According to \cite{Riffonneau2011} the parameters of the STC and NOCT 
measurement conditions are defined as $I_{STC}=1000W/m^2$, $T_{j,STC}=25^{\circ}C$, 
$ I_{NOCT}=800W/m^2$, $T_{amb,NOCT}=20^{\circ}C$, at a wind speed of $1m/s$. The choices 
of other parameters will be given in the evaluation section. However, they are quite
similar to the ones in \cite{Riffonneau2011}.
\subsection{Hot Water Tank}
In our simulation, we modified the hot water tank model, used
in \cite{Laurent1995} and \cite{Chong1979}, in order to keep the complexity low. Specifically, a non-stratifying 
tank model is used, so that the water temperature can be
assumed to be homogeneous.
%In order to keep the \tcb{complexity} of the model low, it is assumed that the water in the tank is always perfectly mixed.
%\tcb{This means} the water has everywhere the same temperature.
Furthermore, the volume of water in the tank is constant. 
This implies that the same amount of water, which is withdrawn 
from the tank is replaced with cold fresh water. In 
addition, the tank walls are not explicitly modeled. Instead 
they are implicitly considered with the constant $a$ 
which influences the speed of heat exchange with the 
surrounding air. Different from the models described 
in \cite{Laurent1995} and \cite{Chong1979}, we assume that the 
electrical heating element
%, which is either electrical or gas driven,
can take an arbitrary power value in the interval $[0,P_{max}]$. 
In other models, 
%it is only possible to switch between zero and maximum power. 
only discrete switching between maximum heating power and off is possible.
Analyzing the thermal energy of the tank leads to the 
differential equation that models the behavior of the water temperature $T(t)$ in the tank as
%The change of heat in the tank is determined by
%\begin{equation}
%\Delta Q_{tank}=-Q_{loss,room}-Q_{loss,demand}+E_{heat-element} \ ,
%\label{eq:tank1}
%\end{equation}
%where $ Q_{loss,room} $ is the loss of heat due to the heat transfer through the tank walls to the ambient air. Refilling the tank with cold water after hot water has been withdrawn causes the loss %of heat $ Q_{loss,demand}$. $ E_{heating}$ is the thermal energy which is added to the tank by the heat element. Taking the derivative with respect to time of Eq.~\eqref{eq:tank1} yields 
%\begin{equation}
%\dot{Q}_{tank}=-\dot{Q}_{loss,room}-\dot{Q}_{loss,demand}+P_{heating} \ .
%\label{eq:tank2}
%\end{equation}
%If the variables are now replaced by their describing functions, the following differential equation for the tank temperature is obtained
%\begin{equation}
%C\frac{d}{dt}T(t)=-a\left(T(t)-T_{room}\right)-c_w W(t) \left(T_{out}-T_{in}\right)+P_{max}u(t) \ 
%\label{eq:tank3}
%\end{equation}
%
\begin{equation}
\begin{split}
\frac{d}{dt}T(t)=-\frac{a}{C}\left(T(t)-\right. & \left. T_{room}\right)-\frac{1}{m_w} W(t)  \\  
&\cdot\left(T_{out}-T_{in}\right) +\frac{P_{max}}{C}u(t) \ .
\label{eq:tank4}
\end{split}
\end{equation}
with $C=c_w m_w$, the thermal capacity of the tank, $a$ a constant describing the heat transfer
trough the tank walls, $c_w$ the specific thermal capacity of water, $m_w$ the mass of water in the 
tank, $W(t)$ the hot water abstraction 
for the desired temperature at time $t$, $T_{out}$ the desired hot water temperature, $T_{in}$ the 
cold water inlet temperature, and $P_{max}$  and $u(t)$ being the design heating power of the heating
element in watts and the normed input to the heating element which can take values between
$\left[0,1\right]$. It is worth mentioning that the specific thermal capacity of water is 
dependent on the water temperature. However, in this simulation, the specific thermal 
capacity of water is assumed to be constant to simplify calculations.
This simplification introduces comparable small errors 
as the temperature at which the tank is operating only varies
within a certain small interval.
%
%Considering the other simplifications in this model, the error %caused by \tcb{doing so is comparably small as
%the temperature at which the tank is operating only vary within %a certain small interval.}
%
%Dividing Eq.~\eqref{eq:tank3} by $C$ leads to the final form of the tank dynamic
%\begin{equation}
%\frac{d}{dt}T(t)=-\frac{a}{C}\left(T(t)-T_{room}\right)-\frac{1}{m_w} W_{flow} \left(T_{out}-T_{in}\right)+\frac{P_{max}}{C}u(t) \ .
%\label{eq:tank4}
%\end{equation}
%
\section{Smart Water Heater Control}
\label{sec:03}
% Ziel? Voraussetzungen/Vorwissen? Algorithmus?
%
The main goal of smart water heater control is to minimize the residents'
total electricity costs by optimizing the control of the temperature in the heated water tank. 
Additionally, the comfort of the residents' should be maintained at all time.
For this purpose, knowledge about the future hot water consumption, produced energy, electricity
demand, and energy prices is required. As these quantities are unknown, an optimization algorithm 
has to rely on predictions of these values. Due to the uncertainty of predictions, it is
necessary to model the unknown future consumption of hot water, produced power,
%energy, 
electricity demand, 
and energy prices as random variables $\omega_k^{w}$, $\omega_k^{PV}$, $\omega_k^{el}$, and 
$\omega_k^{Pr}$, respectively, with $k$ being the current algorithm time step. In our setting, 
we assume that the prediction error is Gaussian
distributed with zero mean and independent for each discrete simulation time step $k$. For example,
the probability density function (PDF) for the random variable of the hot water consumption
$f_{\omega_k^w}(\omega_k^w)$ is $\mathcal{N}(\omega_k^{w,pred},{\sigma_k^{w}}^{2})$ distributed with
$\omega_k^{w,pred}$ as outcome of the hot water consumption prediction. The standard deviation
$\sigma_k^w$ is chosen according to the uncertainty of the particular prediction which is also
assumed to be known. The PDFs of the other random variables 
%are similarly constructed. 
are constructed in a similar way.
Using these random variables and the dynamic of the hot water tank from Eq.~\eqref{eq:tank4}, 
the DP algorithm 
%of \cite{bertsekas1995dynamic} 
seeks an optimal control for a finite horizon in order to minimize the 
energy costs. 

Before the optimization algorithm can be applied, it is necessary to describe the problem
in an appropriate mathematical form. First of all, the continuous time $t$ is 
discretized into equal time steps of length $\Delta t$ and a total simulation time 
length of $N$ steps. Furthermore, the tank temperature $T(k\cdot\Delta t)$ is also discretized and 
in the following represented by the discrete state $x_k$. The normed input of the hot water
heater is also discrete and denoted by $u_k$. 
The system state transition function $f(x_k,u_k,\omega_k)$, required by the DP algorithm,
relates the next state $x_{k+1}$ with the current state $x_k$, input $u_k$, and disturbance 
$\omega_k$. Such a relationship is implicitly given by Eq.~\eqref{eq:tank4}. However, an analytical
solution of this differential equation is not available. Thus, Eq.~\eqref{eq:tank4} is integrated 
over the time step $\Delta t$ using 
%partly
a trapezoidal method and assuming that the hot water consumption and input are constant during
that time step. Finally, the equation is solved for $x_{k+1}$, resulting in an 
approximation of the sought-after explicit relationship
\begin{equation}
\begin{split}
x_{k+1}\!=&rd\Biggl(\!\!\left( 1\!+\!\frac{a\Delta t}{2C}\right)^{\!-1}\!\!\left(\!\left(1-\frac{a\Delta t}{2C}\right)x_k\!+\!\frac{a}{C}T_{room}\Delta t \right.\Biggr.  \\ &
\biggl.\left. -\frac{1}{m_w} \omega_k^w\left(T_{out}-T_{in}\right)\Delta t +\frac{P_{max}}{C}u_k\Delta t \right)\Biggr) \ .
\label{eq:tank5}
\end{split}
\end{equation}
The rounding operator $rd(\cdot)$ in Eq.~\eqref{eq:tank5} is needed due to the discrete modeling of 
the states $x_k$.
% modelling or modeling  AE oder BE? 
%Before DP can be used to minimize the total expected costs 
%
%\begin{equation}
%E\left\{g_N(x_N)+\sum_{k=0}^{N-1}g_k(x_k,u_k,\omega_k) \right\} \, ,
%\label{eq:tank6}
%\end{equation}
%
%which adds the stage costs $g_k(x_k,u_k,\omega_k) $ of every time step $k$, the stage costs have to be %defined.   
%the definition of the the stage costs $g_k(x_k,u_k,\omega_k) $ which arise at time step $k$ for the transition from state is required. 

In the present case, the amount on the energy bill shall be reduced. Hence, an intuitive choice 
for the stage costs is the amount of money which is paid for the exchange of energy with the 
grid during a time step. The exchanged electricity with the grid is the difference between 
produced and consumed electrical energy in the smart home. Thus, the stage costs for one time step are
\begin{equation}
g_k(x_k,u_k,\omega_k) = \left(\left(-\omega_k^{PV}+\omega_k^{el}+P_{max}u_k\right)\Delta t\right)\omega_k^{Pr} \, .
\label{eq:tank7}
\end{equation}
Using the definitions from above the mathematical optimization problem can be formulated. 

Let us denote by $\Pi$ the space of all admissible policies.
The DP algorithm now seeks an optimal policy $\pi^{*}=\{\mu_0^*,\dots,\mu_{N-1}^*\} \in \Pi$  
%the set of admissible policies $\Pi$ 
which minimizes, given a start state $x_0$, the cost-to-go function  
\begin{equation}
\!J_{\pi^{*}}(x_0)\!=\!\min_{\pi \in \Pi} \operatorname*{\mathbb{E}}_{\{\omega_{k}\}}\!\!\left\{g_N(x_N)\!+\!\!\sum_{k=0}^{N-1}g_k(x_k,\mu (x_k),\omega_k) \right\}  \, 
\label{eq:tank8}
\end{equation}
for the considered horizon $N$.
For this purpose, the optimal action $u_k^*=\mu_k^*(x_k)$ for every state at each time step needs
to be calculated. This is achieved by applying the DP algorithm which can be stated according to
\cite{bertsekas1995dynamic} in the following way: The algorithm starts from the terminal
state $x_N$ by defining the costs of the final step as
\begin{equation}
J_{N}(x_N)=g_{N}(x_N) \ .
\label{eq:tank10}
\end{equation}
Afterwards, the algorithm recursively computes the cost-to-go of every state at 
each time step $k=N-1, N-2, \ldots, 1, 0$ using
\begin{equation}
\begin{split}
J_{k}(x_k)=\min_{u_k \in U_k(x_k)} \operatorname*{\mathbb{E}}_{\{\omega_{k}\}} & \Big\{g_k(x_k,u_k,\omega_k) \\
& + J_{k+1}\left(f(x_k,u_k,\omega_k) \right) \Big\} .
\end{split}
\label{eq:tank9}
\end{equation}
The control $u_k$ which minimizes the right side of Eq.~\eqref{eq:tank9} is the optimal 
control $u_k^*$ to apply when the system is in state $x_k$. Thus, minimizing the cost-to-go of 
each state for every time step $k$ will deliver the optimal policy $\pi^*$. In order to solve
Eq.~\eqref{eq:tank9} it is necessary to calculate the included expected value considering the 
contained random variables $\omega_k^{w}$, $\omega_k^{PV}$, $\omega_k^{el}$, and $\omega_k^{Pr} $. 
%For this purpose, we assume that the random variables are independent of each other and as mentioned
%\textcolor{red}{earlier there is either not any interdependency between the random variables of 
%different time steps}. 
For this purpose, we assume that the random variables are both independent of each other and 
independent among different time steps.
Furthermore, we can split the expected value of Eq.~\eqref{eq:tank9} 
by applying the linearity of the expectation operator, i.e.
\begin{equation}
\begin{split}
\operatorname*{\mathbb{E}}_{\{\omega_{k}\}}&\left\{g_k(x_k,u_k,\omega_k)+J_{k+1}\left(f(x_k,u_k,\omega_k) \right) \right\}=\\ & \operatorname*{\mathbb{E}}_{\{\omega_{k}\}}\!\!\left\{g_k(x_k,u_k,\omega_k) \right\} +\!\!\!\!\operatorname*{\mathbb{E}}_{\{\omega_{k}\}}\left\{J_{k+1}\left(f(x_k,u_k,\omega_k) \right) \right\}.
\label{eq:10}
\end{split}
\end{equation}
As the stage costs defined in Eq.~\eqref{eq:tank7} is a linear combination and multiplication 
of random variables assumed to be independent, the following equality holds for the expected 
value of the stage costs
\begin{equation}
\begin{split}
\operatorname*{\mathbb{E}}_{\{\omega_{k}\}}\!\!\left\{g_k(x_k,u_k,\omega_k) \right\}=&\biggl(\!\Bigl(-\operatorname*{\mathbb{E}}_{\{\omega_{k}\}}\{\omega_k^{PV}\}+\operatorname*{\mathbb{E}}_{\{\omega_{k}\}}\{\omega_k^{el}\}\Bigr.\biggr. \\ &\biggl.\Bigl.+P_{max}u_k\Bigr)\Delta t\biggr)\operatorname*{\mathbb{E}}_{\{\omega_{k}\}}\{\omega_k^{Pr}\}  \, .
\label{eq:11}
\end{split}
\end{equation}
In the case of the second summand of Eq.~\eqref{eq:10}, a similar approach 
%to ease the calculation 
is unfortunately not feasible due to the non-linearity of the rounding operator 
in $f(x_k,u_k,\omega_k)$ and the cost-to-go function $J_{k+1}$. As a sequel, the 
expected value has to be determined in a different way as
\begin{equation}
\begin{split}
\!\!\operatorname*{\mathbb{E}}_{\{\omega_{k}\}}\!\!\left\{J_{k+1}\left(f_k(x_k,u_k,\omega_k)\right)\right\} \!= \!\!\int_{-\infty}^{\infty}&  \!\!J_{k+1} \left(f(x_k,u_k,\omega_k^w)\right) \\
& \cdot f_{\omega_k^w}(\omega_k^w)\mathrm{d}\omega_k^w .
\label{exp12}
\end{split}
\end{equation}
However, it is not possible to simply integrate over this expression due to the discrete states
$x_k$ and the rounding operator in $f(x_k,u_k,\omega_k) $. Thus, our method determines first the
likelihood that a certain state $x_{k+1}$ appears. For this purpose, the lower and upper bounds
$(\omega_k^{w,low},\omega_k^{w,up})$ of the hot water consumption interval which leads according 
to Eq.~\eqref{eq:tank5} to state $x_{k+1}$ have to be determined. This is done using 
Eq.~\eqref{eq:tank5} without the rounding operator resulting in a modified function 
$f'(x_k,u_k,\omega_k^{w})$. 
%\marginpar{\textcolor{red}{what is $f'$? It has to be defined.}}
As the upper and lower bounds $(T_k^{up},T_k^{low})$ of the tank 
temperature interval, which is mapped on the discrete state $x_{k+1}$ by the rounding operator, 
are known, these boundaries can be plugged into the modified Eq.~\eqref{eq:tank5} which is solved 
for $\omega_k^{w,low}$ and $\omega_k^{w,up}$, respectively. Considering $x_k$ and $u_k$ as constants 
during one time step the calculation of the lower and upper limit of the hot water 
consumption interval which is mapped on the state $x_{k+1}$ can be stated as
\begin{align}
\omega_k^{w,low}&=f'^{-1}(x_k,u_k,T_k^{up})  \\
\omega_k^{w,up}&=f'^{-1}(x_k,u_k,T_k^{low})	\ ,
\end{align}
with $f'^{-1}$ being the inverted function of $f'$.
Now, as the limits of the interval $\left[\omega_k^{w,low},\omega_k^{w,up}\right]$, which 
is mapped for a particular combination of $x_k$ and $u_k$ to a certain state $x_{k+1}$ are known,
the probability that a state $x_{k+1}$ appears can be calculated as
\begin{equation}
\Pr (x_{k+1})=\int_{\omega_k^{w,low}}^{\omega_k^{w,up}} {f_{\omega_k^{w}}(\omega_k^{w})\mathrm{d}\omega_k^w} \ .
\label{exp14}
\end{equation} 
Using Eq.~\eqref{exp14} it is feasible to rewrite Eq.~\eqref{exp12} in a discretized form as
\begin{equation}
\operatorname*{\mathbb{E}}_{\{\omega_{k}\}}\!\!\left\{J_{k+1}\left(f_k(x_k,u_k,\omega_k)\right)\right\} \approx \!\!\sum_{\forall x_{k+1}}\!\!{J_{k+1}\left(x_{k+1})\right)\Pr (x_{k+1})}.
\label{exp15}
\end{equation}
As a result, the expected value in Eq.~\eqref{eq:tank9} can be computed, and consequently,
a DP algorithm can be successfully applied to the Smart Home. Further details about the
experimental setting of the
Smart Home and the missing variable values are given in the next section.
%In the next section, further details about the 
%experimental setting are given.
%
\section{Experimental setting}
 \label{sec:setting}
\begin{figure*}[ht]
\centering
\includegraphics[width=\textwidth]{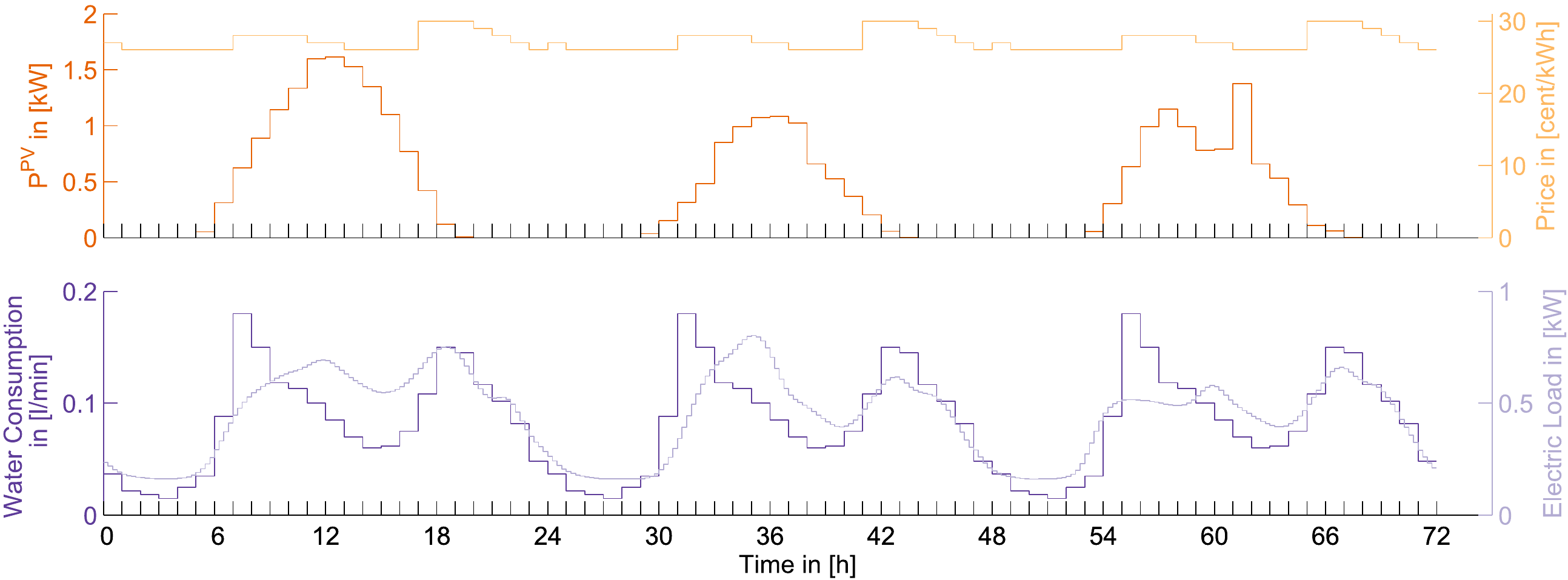}
%\vspace{-1.0em}
\vspace{-0.1cm}
\caption[Prediction Data]{Predicted input data for three days.}
\label{fig:prediction}
\vspace{-0.3cm}
\end{figure*}
In our experiment, we ran the simulation of the water tank along with our control algorithm
over the course of several simulated days. For the sake of readability, in the figures only the
results for a three day timespan are depicted and effects that only occur in the long run are
described alongside if they are of relevance. We assume noisy prediction data to accommodate a 
broad range of forecasting
algorithms. These predictions are generated by assuming consumption and production values for
a typical household and are plotted in Fig.~\ref{fig:prediction}. The PV-generator output is
directly derived using above model together with %from 
natural weather data of the city of Munich. The other values are derived from publicly available 
datasets\footnote{Weather: \url{http://apps1.eere.energy.gov/buildings/energyplus/weatherdata_about.cfm}\\
Load: \url{http://www.ewe-netz.de/strom/1988.php}\\
Prices: \url{https://rrtp.comed.com}\\
Water: Energy Saving Trust, Measurement of domestic hot water consumption in dwellings, 2008}.
Afterwards additive Gaussian noise with a standard deviation of 
$\sigma_k^{w,gen} = \frac{2}{3} \omega_k^{w,pred}$, $\sigma_k^{PV,gen} = \frac{0.5}{3}\omega_k^{PV,pred}$,
$\sigma_k^{el,gen} = \frac{1}{3}\omega_k^{el,pred}$ and $\sigma_k^{Pr,gen} =
\frac{5}{3}\omega_k^{Pr,pred}$ for hot water consumption, PV-generator output, electrical power
consumption and energy price is added to the input data for each experiment, respectively.
Here, variables with a superscript of $gen$ indicate the standard deviations 
%variances 
used in the experiment and
variables with a superscript of $pred$ indicate that this value is taken from the input data at
time step $k$. To compensate for eventual influences of our additive noise, each experiment 
is averaged over 20 independent runs.

To make the solution of the DP problem feasible, continuous time, control input to the water
tank, and temperatures have to be discretized. As we chose a control time interval of 15 $min$, which
is typical 
for a relatively inert system such as 
%\tcp{for a slowly changing system, like }
a water tank, the simulation over three days 
consists of $N = 288$ total timesteps. The water tank temperature is discretized with an interval
of $\Delta x = 0.1 ^{\circ}C$ and the control input with $\Delta u = 0.05$. 
A more limited
classical setting where the heater can only be switched on or off, could be reproduced by
discretizing the control input to two steps only.

In order to prevent the growth of harmful microorganisms, the temperature in the hot water
circulation of a domestic house should always be above $60 ^{\circ}C$. Therefore, we 
%assumed a favourable
%\tcp{set the ideal}
set the
interval for the ideal water tank temperature to be 
in $[T_{min} = 60 ^{\circ}C, T_{max} = 80 ^{\circ}C]$. However, these interval 
bounds are not strictly enforced in the experiment depicted in
Fig.~\ref{fig:Temperature-dynamic} and Fig.~\ref{fig:Temperature-fixed}.

The remaining parameters of the model are summarized in Table~\ref{tab:model-parameters}, where
$N_{PV_s}$ denotes the number of solar panels in series and $N_{PV_p}$ the number of panels connected
in parallel. All model parameters were designed to reflect an average household configuration and
were selected in accordance with \cite{Laurent1995, Riffonneau2011}.
%
% Datengenerierung-->Erstellen der Vorhersagen(woher Daten, welche Modelle werden eventl. verwendet, welche unsicherheiten werden angenommen, wie wird der Fehler genau modelliert, welche unterschiedlichen Größen des Fehlers wird simuliert)
% Simulationszeit und Intervalle
% Welche Ansätze werden verglichen(PI-Regler, shortest path algorithmus, deterministisches Dynamic Programming, unser Ansatz (Dynamic Programming), eventl. noch MPC bzw. andere Ansätze, kein neuronales Netz)
% Was wird verglichen?(welcher Zeitraum,gesamte Energiekosten--->in Abhängigkeit des Optimierungszeitraumes/Zeithorizont + in Abhängigkeit der Genauigkeit der Vorhersagen, Einhaltung des Temperaturfensters)
% Simulationsergebnisse als Durchschnitt der Datensätze 
% Minimale und maximale Temperaturgrenzen in Optimierungsproblemformulierung integrieren? Bzw. hier im experimental setting durch State Space Größe berücksichtigen? 
\begin{table}[htb]
    \label{tab:model-parameters}
    \centering
    \caption[Parameters]{Model Parameters}
        \begin{tabular}[c]{lccc}\toprule
           & Name & Value & Unit \\
            \hline
        PV-         &    $N_{PVs}$   & 5  &  - \\
        Generator   &    $N_{PVp}$    &  2 & -  \\                
                    & $NOCT$      &  45.5 & $^{\circ}C$  \\
                    & $\gamma$    & 0.00043  & $1/^{\circ}C$  \\
                    & $P_{PV,STC}$ & 165 & $W$  \\
            \hline
        Hot         & $a$ & 128.38 & $J/min ^{\circ}C$ \\    
        Water       & $c_w$ & 4.1813 & $J/g ^{\circ}C$ \\
        Tank        & $m_w$ & 196.82 & $kg$ \\
                    & $C$ & $8.22 \cdot 10^5$ & $J/^{\circ}C$\\
                    & $T_{in}$ & 10 & $^{\circ}C$ \\
                    & $T_{out}$ & 60 & $^{\circ}C$ \\
                    & $T_{room}$ & 22 & $^{\circ}C$ \\
                    & $P_{max}$ & 4.5 & $kW $ \\
           % \hline
%        Eigene      & $N$ & 288 & - \\
%        Tabelle?    & $\Delta t$ & 15 & $min$ \\
%        zu Tank?    & $T_{min}$ & 60 & $^{\circ}C$ \\
%        zu Tank?    & $T_{max}$ & 80 & $^{\circ}C$ \\
%        Besser      & $\Delta T$ bzw. $\Delta x$ & 0.1  & $^{\circ}C$  \\ 
%        im          & $\Delta u$ & 0.05  & -   \\
%        Text        &  $u_k$  & $\in [0,\Delta u,\ldots,1]$   &  -  \\
%        ???         &   $x_k$  & $\in [T_{min},\Delta T,\ldots,T_{max}]$   & $^{\circ}C$   \\
%         ?          & $\#$ of Data-Sets & 20 & - \\
%         ?          & $\#$ of days & 3 & - \\
%         ?          & $\sigma_k^w$ & $\frac{2}{3} \omega_k^{w,pred}$ & $kg(l)/min$\\
%         ?used      & $\sigma_k^{w,gen}$ & $\frac{2}{3} \omega_k^{w,pred}$  & $kg(l)/min$\\
%         ?for       & $\sigma_k^{PV,gen}$ & $ \frac{0.5}{3}\omega_k^{PV,pred}$ & $kW$\\
%         ?Data      & $\sigma_k^{el,gen}$ & $ \frac{1}{3}\omega_k^{el,pred}$ & $kW$\\
%         ?generation& $\sigma_k^{Pr,gen}$ & $ \frac{5}{3}\omega_k^{Pr,pred}$ & $cents/kWh$\\
            \bottomrule
        \end{tabular}
        \vspace{-0.2cm}
\end{table}
%
% Stichpunkte:
% -Beschreiben woher man die Vorhersagen der Eingabedaten hat(PV-generator,Temperatur und Irradiance-Daten für München,....) und wie man die Tatsächlichen Kurven kreiert hat.
% (Explizit die Parameter dafür angeben?) 
% - Kurz die Tabelle beschreiben bzw. die Werte plausibel machen (z.B. dadurch, dass man die Werte ähnlich bzw. entsprechend anderer Paper gewählt hat)
% - Beschreiben der Methoden, die zum Vergleich herangezogen wurden (shortestPath, deterministic Dynamic Programming, PI-Regler der konstant die Temperatur hält ---> Parameter % nötig zu erläutern/nennen bzw. quantitativ anzugeben?) <--- wie genau soll man die verglichenen Methoden beschreiben?
% - Beschreiben wie man den Algorithmus implementiert hat 

We compared our stochastic DP algorithm to two other state of the art Dynamic Programming
algorithms. The first is a shortest path version of optimization
for smart home control \cite{Riffonneau2011}. In this variant the discretized water heater temperature 
serves as state in a grid that is spanned on one axis by the time steps and on the other by
the temperature values. %states. 
The algorithm then aims to optimize total cost by 
%\tcp{reverse lookup} %backwards finding
searching backwards, finding
the shortest path through this fully connected grid of states. The second DP algorithm variant 
we compared to, is a deterministic variant, cf. \cite{Tischer2011}. It is similar
to our stochastic algorithm but without explicitly considering the effects of prediction 
and state transition uncertainties. Namely, this algorithm calculates the optimal solution
assuming the real system behaves exactly like the model given the predicted input data 
without noise. 
%Here lies the 
This is a significant difference to our approach in that we can consider possible uncertainties
about the prediction, in particular the hot water consumption, and take them into account for 
further upcoming action scheduling.

A simple Proportional Integral (PI) controller, which is most commonly found in
classical water heaters with a variably controllable heating element, serves as 
baseline algorithm. We do not include other algorithms in the field of optimal heater 
control, as they either do not take the uncertainty of the consumption prediction 
%prediction of consumption
into account, e.g. 
%\cite{zimmerman2011conserve, 
\cite{Pengwei2011} or have to be 
pre-trained extensively involving a tedious parameter tuning, cf. \cite{fuselli2012}.
\section{Results}
 \label{sec:result}
% Plots
% 1.Plot: Eingabedaten: 2 Linien pro Plot, 2mal, breit übereinander
% 2.Plot: Temperature -->RGB-Farben + Style --->gleichbleibend (Titel weglassen) (oben abschneiden)
% 
% 3.Plot:(testen) aufsummieren der Kosten über die Zeit, vergleichen für mit und ohne  
% Verletzungen des Temperaturintervalls als Tabelle (gemittelt über wie viele Runs) (Zeitintervall)
%
% -Temperatur Plots sind nur ein Beispiel für einen Datensatz aus der Menge der 20 generierten Datensätze. 
% -Sollen noch zusätzliche Plots für die tatsächlichen Kurven gemacht werden?
% 
% TODO: Plots:
%    - legende in temperaturplot
%    - farben anpassen
%
\begin{figure}[ht]
\centering
\includegraphics[width=0.45\textwidth]{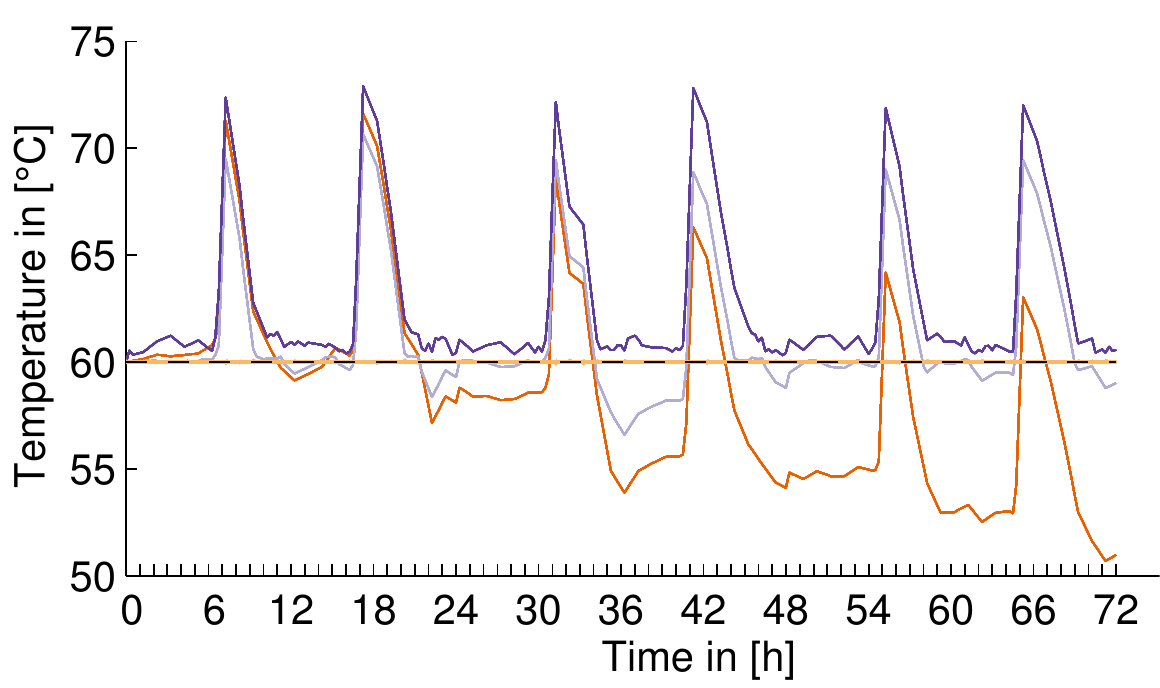}
%\vspace{-0.5em}
\vspace{-0.1cm}
\caption[Temperature]{Simulated water tank temperature for three days (dynamic pricing).}
\label{fig:Temperature-dynamic}
\vspace{-0.3cm}
\end{figure}
\begin{figure}[ht]
\centering
\includegraphics[width=0.45\textwidth]{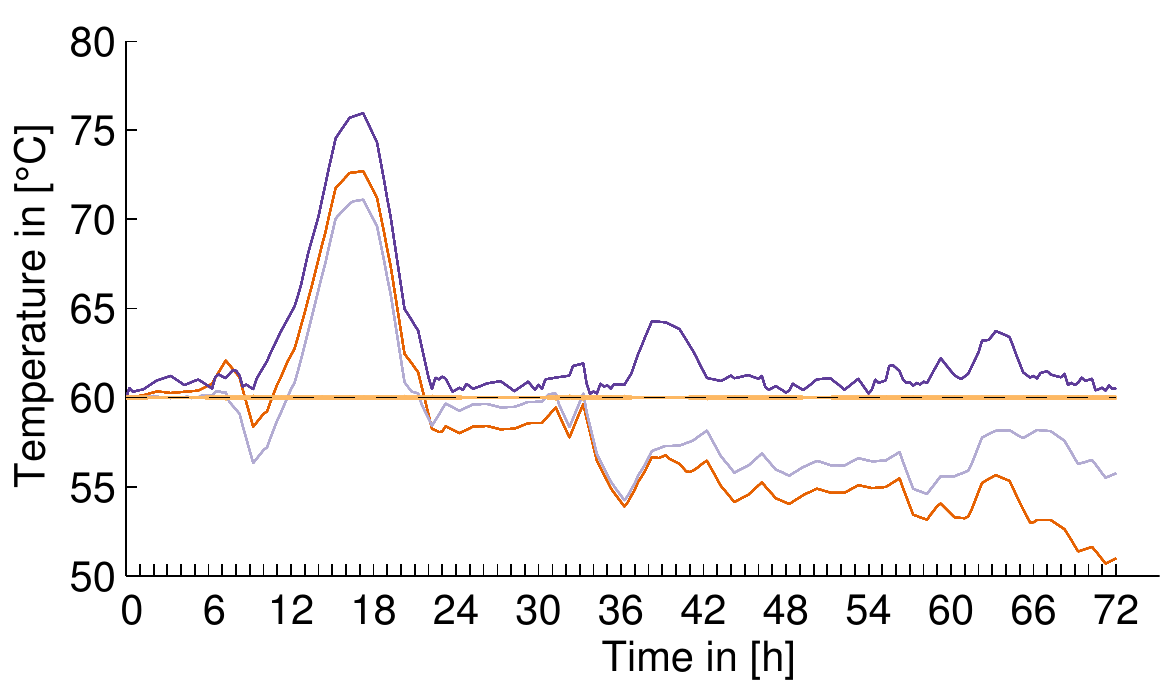}
%\vspace{-0.5em}
\vspace{-0.1cm}
\caption[Temperature]{Simulated water tank temperature for three days (fixed price).}
\label{fig:Temperature-fixed}
\vspace{-0.3cm}
\end{figure}
In Fig.~\ref{fig:Temperature-dynamic} the water temperature for a three-day simulation is depicted
using the dynamic pricing scheme.
%over the course of three days can be observed.
The setpoint of the hot water tank is $60^{\circ}C$ (horizontal dashed line) and it can 
be noticed that the temperature curve of the PI controller (yellow line) is constantly 
holding the desired tank temperature. On the other hand, other algorithms start 
pre-heating the tank in order to compensate for the upcoming hot water consumption.

All DP based method have similar pre-heating behaviour as they all access the same prediction
values. Therefore, at the beginning the behaviour is very similar. As the predictions turn out to be
unreliable, the algorithms that do not consider uncertainty start to decrease in their performance
(shortest path in orange, deterministic DP in light purple). Violations
of the lower temperature bound happen more and more often, while the stochastic DP algorithm 
(dark purple line) can cope
with such unexpected fluctuations. This gap in behaviour amplifies if the runtime of the
experiment is increased. Similar behaviour can be observed, when using a fixed pricing
scheme, as can be seen in Fig.~\ref{fig:Temperature-fixed}. our proposed Stochastic DP
algorithm manages to maintain a
temperature at or above the setpoint, while the other two DP algorithms struggle when
unforeseen water consumption occurs.

In Table~\ref{tab:temperature-violations}, the sum of total violations of temperature bounds
are summed up for the considered three day period. Clearly, the simple shortest path modelling turns
out to be unstable to unforeseen fluctuations as well as deterministic DP. As a low 
water temperature can cause significant inconvenience for the house residents and 
can even make harmful microorganisms to grow, we implemented a compensation mechanism.

\begin{figure}[ht]
\centering
\includegraphics[width=0.45\textwidth]{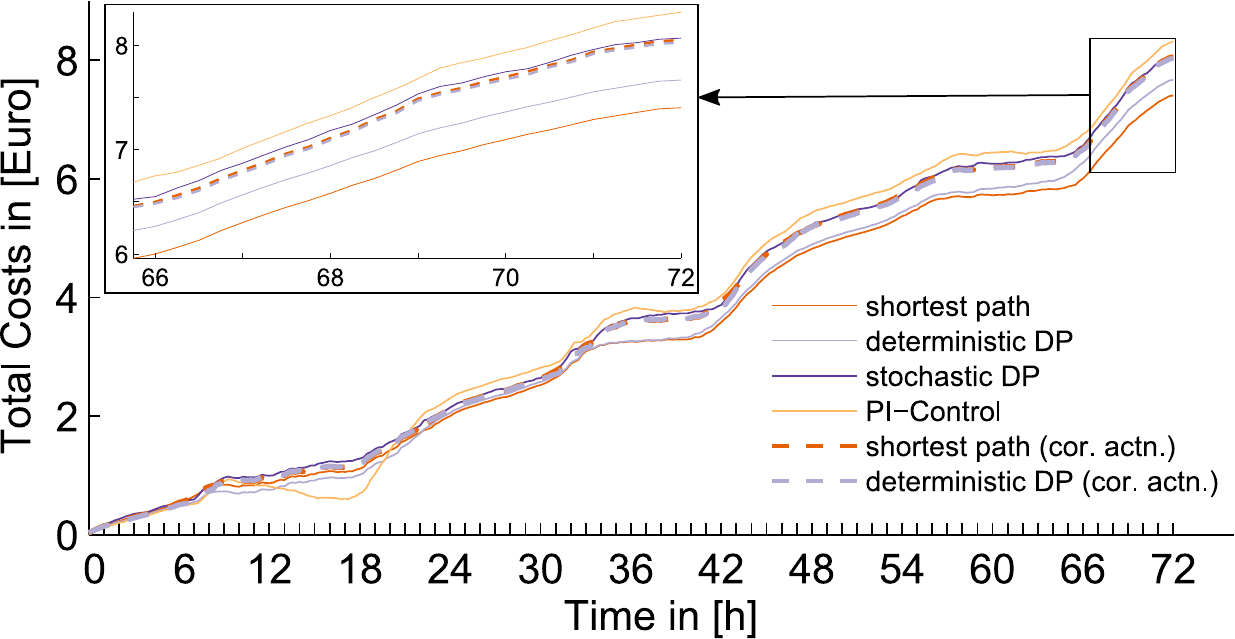}
%\vspace{-0.5em}
%\vspace{-0.5em}
\vspace{-0.3cm}
\caption[Prediction Data]{Accumulated costs graph for a fixed pricing scheme.}
\label{fig:Costs-fix}
\vspace{-0.3cm}
\end{figure}

As soon as the temperature falls below $60^{\circ}C$, the next actions are not determined with
the total costs as an optimization target, but to provide a correction. They are called {\it corrective
actions}. Clearly, those actions increase the total amount of power used. This behaviour can also 
be observed in Fig.~\ref{fig:Costs-fix} for fixed energy prices.
In this graph the total costs that accumulated are plotted. While
for the unmodified algorithms, the two deterministic DP algorithms (solid lines) have the advantage of
using less energy which has to be bought from the grid, at the same time they perform poorly with 
respect to temperature stability. If corrective actions are implemented, the total costs of the 
two deterministic algorithms are more or less the same (dotted lines) as in the stochastic DP case. 
It is to be noted that all three DP algorithms can achieve the goal of increasing eigen consumption
through pre-heating at lower energy costs than the PI controller.
\begin{table}[htb]
    \label{tab:temperature-violations}
    \centering
    \caption{Total Costs and Violation of Temperature Bounds}
        \begin{tabular}[c]{lcccc}\toprule
          Method & \multicolumn{2}{c}{no corrective action}  & \multicolumn{2}{c}{corrective actions}  \\ \cmidrule(lr){2-3} \cmidrule(l){4-5} 
          & Cost/\texteuro &\# of violations & Cost/\texteuro &\# of violations \\
          \hline
         PI-Control         & 9.0755 & - & 9.0755 & - \\
         shortest Path      & 8.0080 & 179 & 8.7484 & 21 \\
         deterministic DP   & 8.6612 & 123 & 8.7631 & 30\\
         stochastic DP      & 8.7712 & 0 & 8.7712 & 0 \\
            \bottomrule
        \end{tabular}
        %\vspace{-0.5cm}
\end{table}
%
% \begin{figure}[ht]
% \centering
% \includegraphics[width=0.50\textwidth]{images/Costs.pdf}
% %\vspace{-0.5em}
% \caption[Prediction Data]{Accumulated costs graph for a dynamic pricing scheme.}
% \label{fig:Costs}
% %\vspace{-1.4cm}
% \end{figure}
%
\section{Conclusions}
 \label{sec:con}
In this paper, we developed a stochastic DP algorithm for controlling water
heater in the scenario of smart home.
Our algorithm demonstrated its promising capability to ensure that the tank
always contains a sufficient amount of hot water to satisfy the demand despite 
the uncertainty
in the warm water consumption prediction. Furthermore, convincing results show the 
feasibility to reduce the energy costs at the same time. The reason for this is 
twofold. Firstly, the eigen consumption is increased,
% First,
% the generated energy is partly consumed by the inhabitants of the smart home
% directly,
enabled by the use of a thermal storage instead of selling to the power grid. Secondly,
the algorithm leverages forecasts of the PV-system output, water consumption, electrical
demand, and energy prices.
%%
%%  extend!!
%
% In the simulation the algorithm demonstrates its ability to control the tank such that 
% it always contains sufficiently warm water to satisfy the demand despite the uncertainty 
% of the warm water consumption prediction. Furthermore, the results are promising in the
% way that they show the feasibility to reduce the costs of the house's energy bill which
% has two reasons. First, in the simulation the generated energy is partly consumed by the 
% inhabitants of the smart home, enabled by the use of a thermal storage, instead of selling 
% the produced energy directly to the grid. Second, the algorithm takes advantage of the 
% forecasts for the PV-system output, warm water consumption, electrical demand, and energy
% prices. 
%
%\section{Acknowledgments}
%
\bibliographystyle{IEEEtran}
\bibliography{IEEEabrv,biblio.bib}
\end{document}